\documentclass{ifacconf}

\pdfoutput=1
\usepackage{natbib}        
\usepackage[hyphens]{url}
\usepackage{marginnote}
\usepackage[dvipsnames]{xcolor}
\usepackage{amsmath}
\usepackage{xfrac}
\usepackage{amssymb}
\usepackage{units}
\usepackage{graphicx}
\usepackage{booktabs}
\usepackage{mathtools}
\renewcommand*{\_}[1]{\ensuremath{_\mathrm{{\scriptsize #1}}}}

\usepackage{verbatim}

\newcommand{\argmin}{\mathop{\mathrm{argmin}}}

\newif\ifextendedversion 
\extendedversiontrue

\newif\ifmargincomments 
\margincommentsfalse

\ifmargincomments
\newcommand{\tmmargin}[2]{{\color{blue}#1}\marginpar{\color{blue}\raggedright\footnotesize [TdM]:\\ #2}}
\newcommand{\obmargin}[2]{{\color{cyan}#1}\marginpar{\color{cyan}\raggedright\footnotesize [OB]:\\ #2}}

\else
\newcommand{\tmmargin}[2]{#1}
\newcommand{\obmargin}[2]{#1}

\fi

\begin{document}
\begin{frontmatter}

\title{Control and Design Optimization of an Electric Vehicle Transmission Using Analytical Modeling Methods\thanksref{footnoteinfo}} 

\thanks[footnoteinfo]{This publication is part of the project NEON (with project number 17628 of the research programme Crossover, which is (partly) financed by the Dutch Research Council (NWO)).}

\author[First]{Olaf Borsboom} 
\author[First]{Thijs de Mooy} 
\author[First]{Mauro Salazar}
\author[First]{Theo Hofman}

\address[First]{Eindhoven University of Technology, 
	5600 MB, Eindhoven, The~Netherlands (e-mail: o.j.t.borsboom@tue.nl, t.mooy@student.tue.nl, \{m.r.u.salazar, t.hofman\}@tue.nl).}
\begin{abstract}                
\obmargin{}{Introductory sentence?}
This paper introduces a framework to systematically optimize the control and design of an electric vehicle transmission, connecting powertrain sizing studies to detailed gearbox design methods.
To this end, we first create analytical models of individual components: gears, shafts, bearings, clutches, and synchronizers.
Second, we construct a transmission by systematically configuring a topology with these components.
Third, we place the composed transmission within a powertrain and vehicle model, and compute the minimum-energy control and design, employing solving algorithms that provide global optimality guarantees.
Finally, we carry out the control and design optimization of a fixed- and two-gear transmission for a compact family electric vehicle, whereby we observe that a two-gear transmission can improve the energy consumption by \unit[0.8]{\%}, while also achieving requirements on gradeability and performance.
We validate our framework by recreating the transmission that is mounted in the benchmark test case vehicle and recomputing the energy consumption over the New European Driving Cycle, where we notice an error in energy consumption of \unit[0.2]{\%}, affirming that our methods are suitable for gear ratio selection in powertrain design optimization.
\end{abstract}


\end{frontmatter}

\section{Introduction}\label{sec:introduction}
In the past decade, we have witnessed a substantial increase in battery-electric vehicle sales~\citep{IEA2021}.
However, the range and the pricing of these vehicles is still a impediment in the large-scale transition, especially when compared to their conventional counterparts~\citep{PaoliGuel2022}.
To improve the performance and cost of these vehicles, we have to investigate the powertrain.
Yet the powertrain is an intricate system, leaving design engineers with numerous choices in topology, technology, sizing, and control.
This problem requires holistic tools to quickly explore the design space, in order to find what powertrain configuration is the right one for the specific vehicle at hand.

\begin{figure}
	\centering
	\includegraphics[width=\columnwidth]{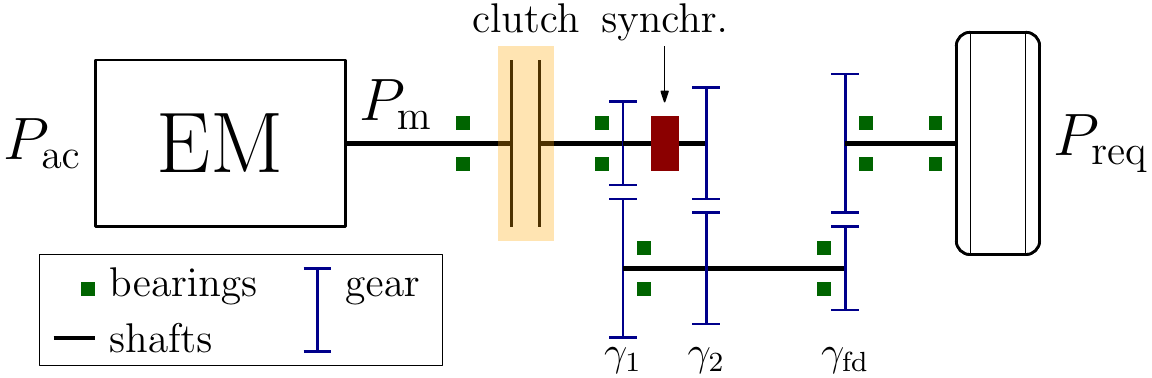}
	\caption{A schematic layout of the powertrain, including an EM and a two-gear transmission. This configuration consists of three shafts, six bearings, three gear pairs, a synchronizer (synchr.), and a clutch.}
	\label{fig:2GTschematic}
\end{figure}

The one component that we address in this paper---the transmission---has been extensively researched for application within conventional vehicle powertrains.
Since in general, an electric motor (EM) can provide sufficient torque over a larger speed range than an internal combustion engine, strictly speaking, electric vehicles do not require a multi-speed transmission, because a fixed-gear transmission (FGT) suffices.
This configuration is also what is most commonly available on the current market~\citep{EVDatabase2022}.
However, given that torque and speed are in fact limited, designers must come to a compromise between acceleration and gradeability on the one hand, and top speed on the other.
For this reason, more high-performance vehicles, from the makes of, e.g., Audi and Porsche, and the Formula E vehicles, have been equipped with multi-speed transmissions~\citep{Koller2021SystematicVehicles, Hewland2022}.
Some vehicle types focusing particularly on energy efficiency have the EMs mounted directly in the wheels, eliminating the need for a transmission altogether.
Determining the suitable transmission for a specific car motivates a methodology that can quickly construct, model, and assess different types of transmissions in the context of a full powertrain.
Configurations like the one in Fig.~\ref{fig:2GTschematic} must be evaluated on their energy efficiency, whilst ensuring a certain performance, in a modular and flexible fashion.

\emph{Related literature:} 
This research relates to three general research streams.
The first stream deals with (hybrid-) electric vehicle powertrain control and design optimization from a system perspective.
This is also known as powertrain \textit{co-design}, and it is typically solved with convex optimization~\citep{PourabdollahEgardtEtAl2018, BorsboomFahdzyanaEtAl2021} or derivative-free solvers~\citep{EbbesenDoenitzEtAl2012,HegazyMierlo2010}.
To ensure computational tractability, these methods either assume a fixed transmission efficiency or disregard losses entirely.

The second stream relates to the transmission design and control optimization at the component level~\citep{QinWuLyu2018, LeiHouEtAl2020, Patil2019}.
However, the optimization problems usually have component-specific objectives, like minimizing the noise and volume or maximizing the strength of the gear teeth, leveraging computationally-expensive finite-element models.
Both the objectives and the methodologies do not connect well to the holistic, system-level perspective that is necessary in powertrain design optimization.

The third and final research stream pertains to more detailed transmission models in powertrain design, striking a balance between accurate modeling and system level design~\citep{MachadoEtAl2021}.
The works in~\cite{HofstetterEtAl2018,AnselmaEtAl2019} minimize the losses, but merely for one configuration.
The procedure in~\cite{KruegerKeinprechtEtAl2022} considers multiple topologies but provides no global optimality guarantees, whereas optimality is guaranteed in~\cite{LeiseEtAl2019}, but fixed efficiencies for all transmission components are considered.

To conclude, to the best of the authors' knowledge, there are no methods that can model and predict the losses of different transmission configurations in a modular and flexible fashion, in the context of a full electric powertrain, optimizing the design and control, whilst guaranteeing global optimality of the solution.

\emph{Statement of contributions:}
To address this issue, this paper presents a modular design and control optimization framework of electric vehicle transmissions.
Specifically, we first derive detailed analytical loss models of all components that compose a transmission: the gears, the shafts, the bearings, the clutches and the synchronizers.
Second, using engineering rules and manufacturer data, we methodically design a transmission system for a powertrain by combining the individual component models in the desired configuration.
Third, we model the full drivetrain and the car, and pose an optimization problem to minimize the energy consumption over a drive cycle.
Finally, we showcase our framework by optimizing the design of the two-gear transmission (2GT) of Fig.~\ref{fig:2GTschematic} and an FGT for a compact family electric car.

\emph{Organization:} 
This paper is organized as follows:
Section~\ref{sec:methodology} presents the transmission component and shifting loss models, the systematic design approach, and the surrounding optimal design and control problem .
We showcase our optimization framework with numerical results in Section~\ref{sec:results}.
Finally, we draw the conclusions in Section~\ref{sec:conclusions}, together with an outlook to future research.

\section{Methodology}\label{sec:methodology}
In this section, we construct the optimization problem, starting with the objective.
Subsequently, we develop a model of the car and the powertrain, whereby we extensively elaborate on the gearbox and its components. 
To increase readability, some equations are purposely omitted from the main text.
\ifextendedversion
    However, the interested reader can find these in the Appendix.
\fi
Moreover, to keep our derivations concise, we will abandon the time dependency $t$, when its obvious from the context.
Finally, we summarize the problem and present the solving method.

\subsection{Objective and Longitudinal Vehicle Dynamics}
The objective of the optimization is to minimize the energy provided to the  EM over a drive cycle:
\begin{equation}
    \min{E\_{ac}},
\end{equation}
where $E\_{ac}$ is the energy consumed at the (electrical) input of the EM.
The car is modeled following a quasi-static approach~\citep{GuzzellaSciarretta2007} in time domain.
The power request at the wheels $P\_{req}$ equals
\begin{equation} \label{eq:wheels}
    P\_{req} = \frac{1}{2} \rho\_{a} c\_d A\_f v^3 + v (m\_v + m\_{gb}) \left( g (c\_r \cos{\beta} + \sin{\beta}) + a \right),
\end{equation}
where $v$, $a$ and $\beta$ are the time-dependent velocity, acceleration, and road inclination, respectively, provided by the drive cycle, $\rho\_a$ is the air density, $c\_d$ is the drag coefficient, $A\_f$ is the frontal area of the car, $m\_v$ is the vehicle mass without the gearbox, $m\_{gb}$ is the gearbox mass, $g$ is the gravitational constant, and $c\_r$ is the rolling resistance coefficient. 

\subsection{Transmission}
In this section, we systematically design a transmission and model the components that a transmission contains, including its losses in an analytical fashion: the spur gear pairs ($\mathrm{g}$), the shafts ($\mathrm{s}$), the bearings ($\mathrm{b}$), the clutches ($\mathrm{cl}$), and the synchronizers ($\mathrm{syn}$).
Using these component models, we can construct multiple transmission configurations in a flexible and modular fashion, and evaluate different transmission types.
In this paper, we focus on the FGT and the 2GT, both designed in two stages.
We describe the modeling procedure for a 2GT, and we can construct an FGT model by eliminating a pair of gears, a clutch, and a synchronizer.

Primarily, we optimize the gear ratio values $\gamma_j$.
These are bounded by
\begin{equation}
    \gamma_j \in [\gamma_\mathrm{min}, \gamma_\mathrm{max}], \quad j \in \{1,2,\mathrm{fd}\}
\end{equation}
where $\gamma_1$ is the ratio of the first gear pair, $\gamma_2$ is the ratio of the second gear pair, $\gamma\_{fd}$ is the final drive ratio, and $\gamma_\mathrm{min}$ and $\gamma_\mathrm{max}$ are the minimum and maximum values of the ratios. 
We ensure that the gear ratios are ordered from a high to a low ratio with the following constraint:
\begin{equation}
    \gamma_1 > \gamma_2.
\end{equation}
Furthermore, the number of gear teeth on the pinions (the driving gears) $N\_t \in \mathbb{N}$ has to be a natural number, as well as the number of teeth on the driven gears $\gamma_j N\_{t}$:
\begin{equation}
    \gamma_j N\_{t} \in \mathbb{N}, \quad j \in \{1,2,\mathrm{fd}\}.
\end{equation} 
We demand the car to be able to launch at the maximum road inclination angle $\beta\_{max}$ in the highest gear ratio $\gamma_1$, guaranteed by the constraint
\begin{equation}
    \gamma\_1 \gamma\_{fd} \geq  \frac{g r\_w m\_v (c\_r \cos{\beta\_{max}} + \sin{\beta\_{max}})}{T\_{m,max}},
\end{equation}
which partially defines our design space, where $T\_{m,max}$ is the maximum output torque of the EM and $r\_w$ is the radius of the wheels.

\subsubsection{Initial Design}
We initialize the design of the gearbox according to the procedures in~\cite{Maciejczyk2011DesignBoxes} and start with the shafts.
In the configuration of the transmission (see Fig.~\ref{fig:2GTschematic}, we consider three shafts ($N\_{s} = 3$): the input shaft ($k=1$), the intermediate shaft ($k=2$), and the output shaft ($k=3$), which are all assumed to be short and therefore infinitely stiff, neglecting shaft torsion. 
The torque limit of the gearbox is then determined by the radius of the shafts. 
We determine the radius of a shaft $r_{\mathrm{s},k}$ using the following equation:
\begin{equation}
    r_{\mathrm{s},k} = \sigma\_s T_{k,\mathrm{max}}^{\frac{1}{3}}, \quad k \in \{1, 2, 3\}
\end{equation}
where $T_{k,\mathrm{max}}$ is the maximum torque that shaft $k$ needs to cope with, defined by $T\_{m,max}$ and the gear ratio $\gamma_j$, and
\ifextendedversion
     $\sigma\_s$ is a parameter defined in the Appendix.
\fi

The pinions of the first gear pair and final drive are designed as the shaft with teeth mounted on the circumference, which means that the root radius of these pinions is equal to the radius of the shaft. 
Their pitch radius $r_{\mathrm{p},j}$ is then determined using
\begin{equation}
    r_{\mathrm{p},j} = \frac{r\_{s,1}}{1-\frac{\sigma\_{p}}{N\_t}}, \quad j \in \{1,\mathrm{fd}\}, \quad k \in \{1,2\},
\end{equation}
\ifextendedversion
    where $\sigma\_{p}$ is a pitch radius parameter defined in the Appendix.
\fi 
In general, the radius of a driven gear $r_{\mathrm{g},j}$ is equal to
\begin{equation}
    r_{\mathrm{g},j} = r_{\mathrm{p},j} \gamma_j, \quad j \in \{1, 2, \mathrm{fd}\}.
\end{equation}
For every gear pair, except the final drive, the distance between the centers of the pinion and gear $d\_g$ has to be equal, since they must be mounted on the same shafts:
\begin{equation}
    d\_g = r\_{p,1} + r\_{g,1} = r\_{p,2} + r\_{g,2}.
\end{equation}
In our framework, we determine the value of $d\_g$ from the design of the first gear and comply with that distance in the second gear.
Since the pitch radii of the first gear pair are now known, the pitch radius of the pinion for the remaining gear pair $r_{\mathrm{p},2}$ is equal to
\begin{equation}
    r_{\mathrm{p},2} = d\_{g}\left(1-\frac{1}{1+\gamma_2}\right).
\end{equation} 
We calculate the outer radius of the clutch $r\_{cl,o}$ using the methods in~\citep{Bak2021TorqueConnections}, namely
\begin{equation}
    r\_{cl,o} = \sigma\_{cl,o} T\_{m,max}^{\frac{1}{3}},
\end{equation}
\ifextendedversion
    where $\sigma\_{cl,o}$ is a clutch radius parameter defined in the Appendix.
\fi
The inner radius of the clutch $r_\mathrm{cl,i}$ then becomes
\begin{equation}
    r\_{cl,i} = \phi\_{cl}r\_{cl,o},
\end{equation}
where $\phi\_{cl}$ is the predefined ratio between the outer and inner radius of the clutch.
The force required to engage the clutch plates $F\_{cl}$ is equal to
\begin{equation}
    F\_{cl} = \sigma\_{cl} \left(r\_{cl,o}^2 - r\_{cl,i}^2\right),
\end{equation}
\ifextendedversion
    where $\sigma\_{cl}$ is a clutch force parameter defined in the Appendix.
\fi

Since the size of the rotating part of a bearing is small, we neglect its inertia and we consider its mass $m\_b$ to be constant.
The latter can be found in manufacturer catalogues, including its width $b\_b$~\citep{SKFGroup2018RollingBearings}.
The inertia and mass of all remaining modeled gearbox components are approximated by assuming all components are thick-walled cylindrical tubes with open ends. 
Therefore, the mass of a component $m_c$ can be calculated using
\begin{equation}
    m_c = \rho_c b_c \pi(r_{c,\mathrm{o}}^2 - r_{c,\mathrm{i}}^2), \quad c \in \{\mathrm{g}, \mathrm{s}, \mathrm{cl}, \mathrm{syn}\},
\end{equation}
where $\rho_c$ is the density of the material of the component, $b_c$ is the width ($c \in \{\mathrm{g}, \mathrm{cl}, \mathrm{syn}$\}), or length ($c = \mathrm{s)}$ of the component, and $r_{c,\mathrm{o}}$ and $r_{c,\mathrm{i}}$ are the outer and inner radius, respectively.
In our case, $r\_{s,i}=0$ and $r\_{g,i} = r\_{syn,i} = r\_{s,o}$, since we consider the gears and synchronizers to be directly mounted on the shafts. 
We assume $b\_g$ is related to the gear radius
\ifextendedversion
(see Appendix),
\fi
which is equal to $b\_{syn}$.
The length of a shaft $b\_s$ is the summation of the width of all the components mounted on it.
This makes the inertia $J_c$ equal to
\begin{equation}
    J_c = \frac{1}{2} m_c (r_{c,\mathrm{o}}^2 + r_{c,\mathrm{i}}^2), \quad c \in \{\mathrm{g}, \mathrm{s}, \mathrm{cl}, \mathrm{syn}\}.
\end{equation}
For the complete gearbox, the mass $m\_{gb}$ and inertia $J\_{gb}$ are equal to the summation of all the masses and inertias of the components, respectively.

\subsubsection{Gear losses}
The gear losses are split into four types: 
The sliding ($\mathrm{sl}$) and roll ($\mathrm{r}$) friction, and the oil churning ($\mathrm{ch}$) and windage ($\mathrm{wi}$) losses. 
The sliding and roll friction losses are modeled using the elasto-hydrodynamic lubrication (EHL) model \citep{Xu2007PredictionPairsb}. 
The sliding friction loss is equal to
\begin{equation}
    F\_{g,sl} = \mu W,
\end{equation}
\ifextendedversion
    where $\mu$ is the sliding coefficient fully defined in the Appendix
\fi
and $W$ is the normal force on the tooth.
By calculating the sliding coefficient for a slide-to-roll ratio for a gear mesh cycle, the average sliding friction during rotation can be determined.

\ifextendedversion
The expression of the rolling friction loss $F\_{g,r}$ is given in the Appendix.
\fi

As the gears are only partially submerged in lubrication oil, there are both oil churning losses (dragging through the oil) and windage losses (moving through the air).
We use the same model for approximating both these types of losses, but with different material properties, expressed in three terms~\citep{Seetharaman2009}:
The losses at the circumference $P\_{ch,p}$, the faces $P\_{ch,f}$, and teeth cavities $P\_{ch,c}$. 
\ifextendedversion
    The full expressions of these models are given in the Appendix.
\fi 

The total power lost due to oil churning $P\_{g,ch}$ is then equal to
\begin{equation}\label{eq:g,ch}
    P\_{g,ch} = P\_{ch,p} + P\_{ch,f} + P\_{ch,c},
\end{equation}
using the parameters for oil, and the windage power loss can be expressed as
\begin{equation}
    P\_{g,wi} = P\_{ch,p} + P\_{ch,f},
\label{eq:g,wi}
\end{equation}
using the parameters for air.
We do not include the teeth cavity losses in the windage model, since these losses are negligible according to~\cite{Seetharaman2009}.
The total power loss caused by the gears $P\_g$ is summed as
\begin{equation}
    P\_g = (F\_{g,sl} + F\_{g,r})v\_e + P\_{g,ch} + P\_{g,wi},
\end{equation}
where $v\_e$ is the speed of the contact surface. 

\subsubsection{Bearings}
The bearings are modeled by dividing the losses into four terms: 
The moments of rolling ($\mathrm{r}$), sliding ($\mathrm{sl}$), seal ($\mathrm{s}$), and drag friction ($\mathrm{drag}$)~\citep{SKFGroup2008TheMoment}. 
The bearing rolling and sliding frictional moments are similar to the gear counterparts. 
We disregard the seal frictional moment, since we assume that the gearbox is a closed system and does not require sealing.
The drag friction is a combination of the remaining losses, i.e., oil churning and windage.

The rolling frictional moment $T\_{b,r}$ is equal to
\begin{equation}
    T\_{b,r} = \sigma\_{b,r} \omega\_{b}^{0.6}
\end{equation}
\ifextendedversion
    where $\sigma\_{b,r}$ is a bearing roll parameter defined in the Appendix
\fi
and $\omega\_{b}$ is the rotational speed of the bearing.

The sliding frictional moment $T\_{b,sl}$ is \obmargin{represented}{@MaS how can we simplify this?} by
\begin{equation}
    T\_{b,sl} = G\_{sl}(\phi\_{bl}\mu\_{bl} + (1-\phi\_{bl})\mu\_{EHL}),
\end{equation}
where $G\_{sl}$ is a bearing and load-dependent variable, $\phi\_{bl}$ is a weighting factor, $\mu\_{bl}$ is a constant depending on movement (stationary or rotating), and $\mu\_{EHL}$ is the sliding friction in full-film EHL conditions.

The drag frictional moment $T\_{b,drag}$ is calculated using
\begin{equation}
    T\_{b,drag} = \sigma\_{b,drag} \omega\_b^2 + \bar{\sigma}\_{b,drag} \omega\_b^{\epsilon\_{b,drag}},
\end{equation}
\ifextendedversion
    where $\sigma\_{b,drag}$, $\bar{\sigma}\_{b,drag}$, and $\epsilon\_{b,drag}$ are all bearing drag parameters defined in the Appendix.
\fi

To summarize, the total bearing power loss $P\_b$ is equal to
\begin{equation}
    P\_b = (T\_{b,r} + T\_{b,s} + T\_{b,drag})\omega\_b,
\end{equation}
whereby we assume to have two bearings per shaft in our configuration.

\subsubsection{Shafts}
Depending on the location of the shaft in the transmission system, it is possibly (partially) submerged by oil.
Therefore we can model the losses incurred by the shafts $P\_s$ in one term, namely oil churning.
The expression is equal to the one used for the churning of the gears, and is given in \eqref{eq:g,ch}.

\subsubsection{Clutches}
We equip the powertrain with a dry clutch, which leaves us with windage losses as the mere loss term for the clutch losses $P\_{cl}$.
Its model is identical to~\eqref{eq:g,wi}.

\subsubsection{Synchronizers}
The oil layer between the synchronizers and the gears causes a drag power when there is a difference in velocity between the synchronizer and gear, i.e., when a gear pair is disconnected.
This drag power $P\_{syn}$ is equal to~\citep{Dogan2001}
\begin{equation}
    P\_{syn} = \sigma\_{syn}(\omega\_{syn}-\omega\_g)\omega\_{syn},
    \label{eq:syn_1}
\end{equation}
\ifextendedversion
    where $\sigma\_{syn}$ is a synchronizer drag parameter defined in the Appendix, $\omega\_{syn}$ is the speed of the synchronizer (equal to the speed of the previously selected gear), and $\omega\_g$ is the speed of the currently selected gear.
\fi

\subsubsection{Full Gearbox Model}
As mentioned previously, we model the powertrain using the quasi-static (backward-facing) modeling approach, to make fair comparisons between configurations.
However, all transmission loss models introduced so far are forward-facing.
We overcome this issue by first establishing a loss map for the operating range of each gear in the transmission, and subsequently simulating the drive cycle in a quasi-static fashion and interpolating the losses in the simulation.
We minimize the interpolation errors by creating the loss maps with a fine grid.
The combined power losses $P\_{gb,loss}$ are equal to
\begin{multline}
    P\_{gb,loss} = P\_{g,fd} +  \sum_{k = 1}^{N\_{s}} \left(2P_{\mathrm{b},k} + P_{\mathrm{s},k}\right) + J \frac{va}{r\_w^2} \\
    \begin{cases}
        + P\_{g,1} & \text{ if } N\_g = 1 \\
        + P\_{g,1} + P\_{g,2} + P\_{cl} + P\_{syn} & \text{ if } N\_g = 2,
    \end{cases}
\end{multline}
where $J$ is the inertia of the complete drivetrain including the wheels.

\subsubsection{Shifting}
We model the actuators that engage the clutch and synchronizers during shifting with a fixed efficiency $\eta_{\mathrm{a},c}$.
The energy consumed by an actuator $E_{\mathrm{a},c}$ is then equal to
\begin{equation}
    E_{\mathrm{a},c} = \frac{2F_{\mathrm{a},c}h_{\mathrm{a},c}}{\eta_{\mathrm{a},c}}, \quad c \in \{\mathrm{cl}, \mathrm{syn}\}
    \label{eq:a}
\end{equation}
where $F_{\mathrm{a},c}$ is the force needed by the actuator to move the clutch or synchronizer, and $h_{\mathrm{a},c}$ is the distance over which the clutch or synchronizer is moved.
Specifically, the actuation distance of a synchronizer is assumed to be related to its width, $h_{\mathrm{a,syn}} = \frac{b\_{syn}}{2}$.

The synchronization of the clutch plates is approximated through the slip speed $\omega\_{sl}(t)$, which is equal to~\citep{Park2021OptimizationPowertrainb}
\begin{equation}
    \omega\_{sl}(t) = \mathit{\Delta}\omega\_{cl,0}e^{-\zeta t},
    \label{eq:cl_engage}
\end{equation}
where $\mathit{\Delta}\omega\_{cl,0}$ is the initial relative angular velocity between the two clutch plates, i.e. between the EM speed and the speed of the wheels through the currently selected gear, $\zeta$ is the rate at which the pressure in the clutch increases --- and therefore dependent on the clutch springs and actuator ---, and $t$ is the time. 
We match the speeds up to a tolerance equal to $\omega\_{sl}\leq$ \unit[1]{rad/s}.
The energy lost during this shifting phase $E\_{sh}$, including the inertial losses, is equal to
\begin{multline}
        \mathit{\Delta}E\_{sh} = \int_0^{t\_{sh}}\left((J_1+J_2)T\_{m}+J_1J_2\zeta\omega\_{sl}(t)\right)\frac{\omega\_{sl}(t)}{J_1+J_2}\mathrm{d}t\\
        + E\_{a,cl} + E\_{a,syn},
\end{multline}
where $t\_{sh}$ is the duration of a gear shift, $J_1$ is the inertia of the components between the EM and first clutch plate, and $J_2$ is the inertia of the components between the second clutch plate and the wheels. 

\subsection{Electric Motor}
For any negative power requests, the EM power output is saturated to the maximum EM power $P\_{m,max}$. 
The EM power output $P\_{m}$ then equals
\begin{equation}
P\_{m} =
    \begin{cases}
        P\_{req} + P\_{gb,loss} & \text{ if } P\_{req} \geq 0 \\        
        \max{\left(-P\_{m,max},  r\_{b}P\_{req} + P\_{gb,loss} \right)} & \text{ if } P\_{req} < 0,     
    \end{cases}
\end{equation}
where $r\_b$ is the regenerative braking fraction. 
The speed of the EM $\omega\_{m}$ is equal to
\begin{equation}\label{eq:omega_m}
    \omega\_{m} = \gamma_{j}\gamma\_{fd}\frac{v}{r\_w}, \quad j \in \{1,2\},
\end{equation}
which is equivalent to the input speed of the gearbox, where $j$ is the selected gear.
We model the EM using a quasi-static efficiency map:
\begin{equation}\label{eq:em_effmap}
    P\_{ac} =
    \begin{cases}
    \frac{P\_m}{\eta\_m(P\_{m},\omega\_m)} & \text{ if } P\_m \geq 0 \\
    \eta\_m(P\_{m},\omega\_m)P\_m & \text{ if } P\_m < 0.
\end{cases}
\end{equation}

\subsection{Optimization Problem}\label{subsec:optimization}
In this section, we summarize the optimal design and control problem of the 2GT. 
Due to space limitations, we omit the FGT problem, considering it is a simplified, design-only instance of the 2GT problem.
The state variables are $x = (E\_{ac}, i)$, where $i\in \{1,2\}$ is the current gear.
The design variables are $p = \gamma_j$ where $j \in \{1,2,\mathrm{fd}\}$.
Finally, the control variable is $u \in \mathcal{U}(i)$, which is equal to the shift input:
$\mathcal{U}(1) \in \{0, 1\}$ (only an upshift or no shifting is possible), otherwise, $\mathcal{U}(2) \in \{-1, 0\}$ (merely a downshift or no shifting).
The design problem of an FGT is identical, with simplifications in the sense that $i= 1$, $j \in \{1,\mathrm{fd}\}$, and no control input is present.
\obmargin{}{@MaS: how can I express the shift differential equation in continuous time?}
\begin{prob}[Optimal Co-Design Problem] \label{prob:main}
The 2GT minimum energy design and control are the solution of
\begin{equation*}
\begin{aligned}
&\!\min_{u,p} & &E\_{ac}(p, u) = \int_0^{t\_f} P\_{ac} \mathrm{d}t + \mathit{\Delta} E\_{sh}(u)\\
& \textnormal{s.t. } & & i\_{next} = i\_{current} + u  \\
& & & \eqref{eq:wheels}-\eqref{eq:em_effmap}.
\end{aligned}
\end{equation*}
\end{prob}

\begin{figure}
	\centering
	\includegraphics[width=0.4\columnwidth]{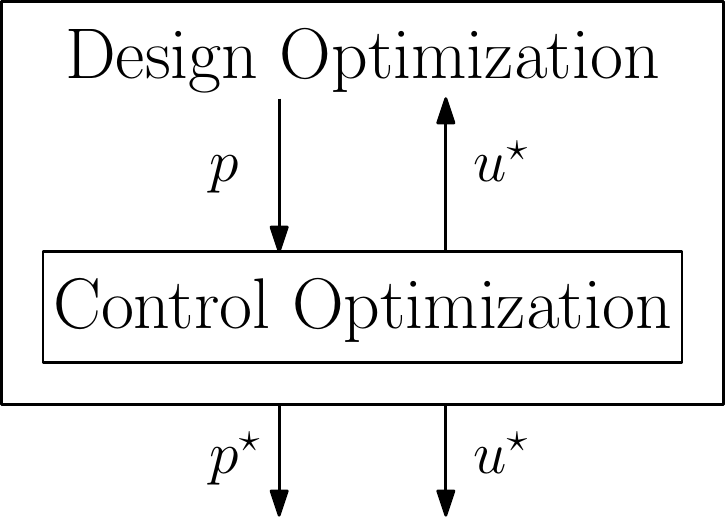}
	\caption{A visualization of the nested solution strategy of the 2GT design and control optimization problem.}
	\label{fig:problem}
\end{figure}

We reformulate this problem to an equivalent nested co-design problem, which is visualized in Fig.~\ref{fig:problem}~\citep{Fathy2003}.

\begin{prob}[Nested Co-Design Problem] \label{prob:reformulated}
The 2GT minimum energy design and control are the solution of the nested problem
\begin{equation*}
\begin{aligned}
&\!\min_{p} & &E\_{ac}(p, u^{\star}(p)) = \int_0^{t\_f} P\_{ac} \mathrm{d}t + \mathit{\Delta} E\_{sh}(u^{\star}(p))\\
& \textnormal{s.t. } & & u^{\star}(p) = \argmin_u E\_{ac}(p, u) \\
& & & \eqref{eq:wheels}-\eqref{eq:em_effmap}.
\end{aligned}
\end{equation*}
\end{prob}
\obmargin{}{doesn't add much maybe}
\subsection{Discussion}\label{subsec:discussion}
A few comments are in order.
First of all, a number of assumptions, approximations, and generalizations are made in the configuration, the design process, and component selection.
Although these assumptions are based on industrial standards and examples, our framework can be modified to other desired initial designs and configurations.
Second, we do not incorporate dynamic thermal effects, since gearbox \obmargin{temperatures are generally constant in steady-state conditions and do not affect the design.}{Not sure}
Third, we also do not consider volumetric constraints, yet these can be taken into consideration when data is available of the specific car.
Moreover, this framework serves as an instrument for gear ratio selection in initial stages of powertrain design, where packaging issues can be addressed in later phases.
Finally, we only consider two ratio stages --- first (and second) gear and the final drive --- in a spur-gear fashion.
Usually, a drivetrain also contains a differential, and can contain other types of gearing, like planetary gear sets or helical gears, etc. 
This can be implemented in our framework with additional modeling, which will be left to \obmargin{a future extension of this paper.}{@MaS, what do you think are the most important points?}
\section{Results}\label{sec:results}
In this section, we present the numerical results of applying the systematic modeling methods and the co-design optimization framework presented in Section~\ref{sec:methodology} to optimize the design of an FGT and the control and design of a 2GT in a compact family electric vehicle.
We select the vehicle and baseline powertrain parameter values of a Nissan Leaf, shown in Table~\ref{tb:vehicle_powertrain}~\citep{Carinf2022}, and the Common Artemis Drive Cycle (CADC) as the driving mission, since it is a relatively challenging cycle.
The most relevant gearbox modeling parameter values are given in Table~\ref{tb:gearbox}, 
\ifextendedversion
    while the more specific parameters are defined in Table~\ref{tb:appendixparams} in the Appendix.
\fi 
We discretize the problem using the Euler Forward method, with a sampling time of \unit[1]{s}.
Since the optimization problem is highly non-linear, we solve the design problem in the outer layer with an exhaustive search, while the control problem in the inner layer is solved using dynamic programming.
In other words, both of these layers are solved using methods offering global optimality guarantees, providing us with a solution with the same properties.
Since the number of teeth on any gear must be a natural number, we discretize the design space with gear step size $\Delta\gamma_j = \frac{1}{N\_t}$.
The design space is further reduced by only considering designs that can attain the maximum speed and acceleration requested by the CADC, $v\_{max}$ and $a\_{max}$, respectively, and the road inclination demand of $\beta\_{max}$.
This leaves us with 355 feasible designs for the FGT, and 65820 for the 2GT.
All computations were performed on an AMD Ryzen 5 5600x CPU and NVIDIA GeForce RTX 3070 Ti GPU, resulting in computation times of around \unit[40]{s} for the FGT and around \unit[7]{h} for the 2GT.

\begin{table}
\begin{center}
\caption{Vehicle \& EM Parameters}\label{tb:vehicle_powertrain}
\begin{tabular}{llll}
\toprule 
\textbf{Symbol} & \textbf{Variable} & \textbf{Value} & \textbf{Unit} \\ \hline
\multicolumn{4}{c}{\textit{Vehicle}} \\
$A\_{f}$ & Frontal Area & 2.27 & [\unit{m$^2$}]\\ 
$c\_{d}$ & Drag coefficient & 0.29 & [1]\\
$c\_r$ & Rolling resistance coefficient & 0.01 & [1]\\
$g$ & Gravitational constant & 9.81 & [\unit{N/kg}]\\
$\rho_{a}$ & Air density & 1.225 & [\unit{kg/m$^3$}]\\
$m\_{v}$ & Vehicle mass without gearbox & 1513 & [\unit{kg}]\\
$m\_{w}$ & Wheel mass & 20 & [\unit{kg}]\\
$r\_{w}$ & Wheel radius & 0.327 & [\unit{m}]\\
$r\_{b}$ & Regenerative brake fraction & 0.55 & [1]\\

\midrule
\multicolumn{4}{c}{\textit{Electric Motor}} \\ 
$T\_{m,max}$ & Maximum EM output torque & 287 & [\unit{Nm}]\\
$\omega\_{m,max}$ & Maximum EM speed & 10,000 & [\unit{rpm}] \\
\end{tabular}
\end{center}
\end{table}

\begin{table}
\begin{center}
\caption{Gearbox Parameters}\label{tb:gearbox}
\begin{tabular}{llll}
\toprule 
\textbf{Symbol} & \textbf{Variable} & \textbf{Value} & \textbf{Unit} \\ \hline
$N\_t$ & Number of teeth on a pinion & 17 & [1]\\
$\rho_c$ & Density of steel & 7,850 & [\unit{kg/m$^3$}]\\
$\beta\_{max}$ & Maximum road inclination & 7.5 & [$^\circ$]\\
$\gamma\_{min}$ & Minimum gear stage ratio & 1 & [1]\\
$\gamma\_{max}$ & Maximum gear stage ratio & 5 & [1]\\
$b\_{cl}$ &  Clutch width & 16 & [\unit{mm}]\\
$b\_{s,3}$ &  Output shaft length & 1.536 & [\unit{m}]\\
$t\_{sh}$ & Shifting time & 0.5 & [\unit{s}]\\
$h\_{a,cl}$ & Clutch actuation distance & 4 & [\unit{mm}]\\
$F\_{a,syn}$ & Synchronizer actuation force & 300 & [\unit{N}]\\
$\eta_{\mathrm{a},c}$ & Actuator efficiency & 0.8 & [1]\\
$\phi\_{cl}$ & Clutch radius ratio & 0.7 & [1]\\

\end{tabular}
\end{center}
\end{table}

\subsection{Numerical Results}
\begin{figure}
    \centering
    \includegraphics[width=\columnwidth]{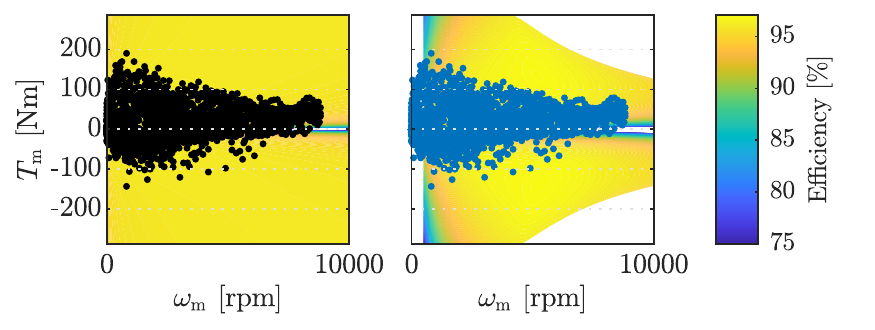}
    \caption{The efficiency maps of the optimally-designed FGT (left) and the EM (right) with the gradeability constraint, including the operating points over the CADC.}
    \label{fig:Leaf}
\end{figure}

\begin{figure}
    \centering
    \includegraphics[width=\columnwidth]{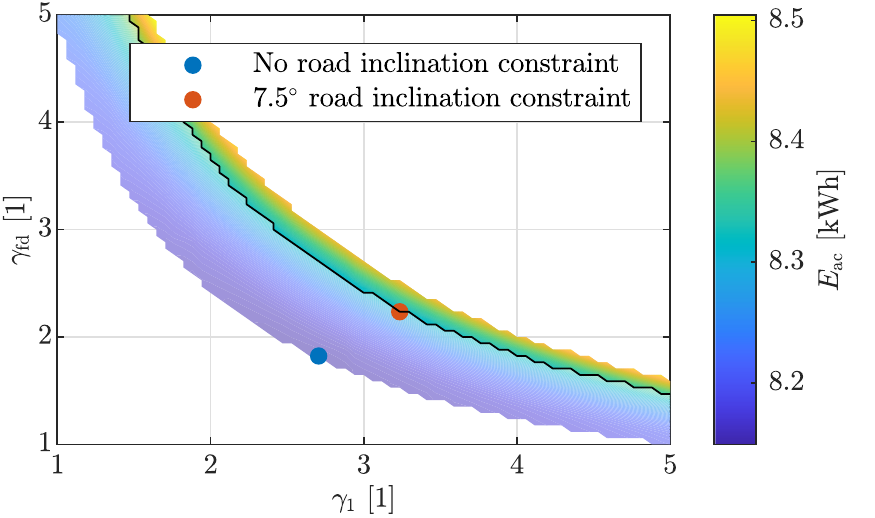}
    \caption{The optima and sensitivity of the optimal control solutions in the feasible design set of the FGT, excluding the gradeabilty constraint (the full set) and including it (the set above the black boundary).}
    \label{fig:Leaf_exhaustivesearch}
\end{figure}

The solutions are summarized in Table~\ref{tb:transmission_designs}.
We solve the FGT design problem for two cases: with and without a road inclination constraint. 
The obtained solutions are shown in Fig.~\ref{fig:Leaf_exhaustivesearch}, where they are framed within the objective values of the full design space.
The test case with a road inclination constraint $\beta\_{max} = 7.5^\circ$ results in a design of $\gamma_1 \approx 3.2$, $\gamma\_{fd} \approx 2.2$.
The subsequent efficiency maps of both gear stages combined, and the EM are shown in Fig.~\ref{fig:Leaf}, together with the interrelated operating points over the CADC.
The total energy consumption at the end of the cycle is equal to $E\_{ac}(t\_f) =$ \unit[8.30]{kWh}.
When we ignore the gradeability constraint, we obtain a design of $\gamma_1 \approx 2.7$, $\gamma\_{fd} \approx 1.8$, and $E\_{ac}(t\_f) =$ \unit[8.12]{kWh}; a difference of \unit[2.2]{\%}.
However, the maximum road inclination the vehicle can handle with this gearbox design is only $5^\circ$.

\begin{figure}
    \centering
    \includegraphics[width=\columnwidth]{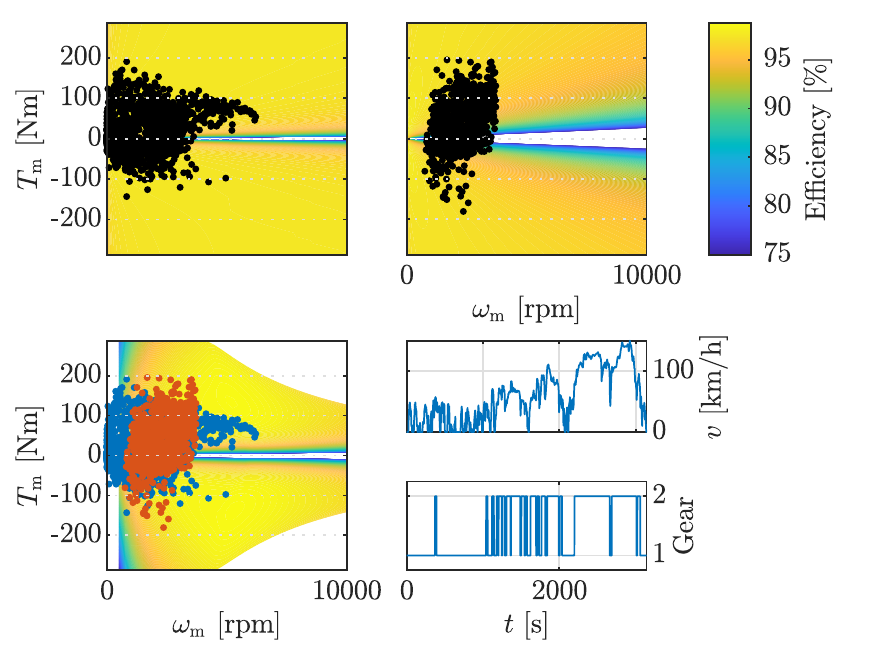}
    \caption{The efficiency maps of the optimally designed 2GT (first gear: top left, second gear: top right) and the EM (bottom left), including the operating points, and the shifting trajectory.}
    \label{fig:Leaf_2spd}
\end{figure}

If we solve the 2GT design and control problem for the same vehicle, we arrive at the solutions shown in Fig.~\ref{fig:Leaf_2spd} and Table~\ref{tb:transmission_designs}.
This is a design of $\gamma_1 \approx 2.82$, $\gamma_2 \approx 1.18$, $\gamma\_{fd} \approx 2.59$, and $E\_{ac}(t\_f) =$ \unit[8.23]{kWh}; an energy consumption decrease of \unit[0.8]{\%} with respect to the FGT solution with the gradeability constraint, since both the transmission and EM are allowed to operate in more efficient regions. 
We observe a reasonable shifting trajectory, where the second gear is selected only in higher-speed regions.

\begin{table}
\begin{center}
\caption{Transmission Design Solutions}\label{tb:transmission_designs}
\begin{tabular}{lllll}
\toprule 
\textbf{Configuration} & ${\gamma_1}$ & ${\gamma_2}$ & ${\gamma\_{fd}}$ & ${E\_{ac}(t\_f)}$\\ \hline
\rule[1ex]{0pt}{1ex}FGT (Baseline) & $\frac{31}{17}$ & - & $\frac{74}{17}$ & \unit[8.48]{kWh} \\ [1ex]
FGT (Gradeab. constr.) & $\frac{55}{17}$ & - & $\frac{38}{17}$ & \unit[8.30]{kWh} \\ [1ex]
FGT (No gradeab. constr.) & $\frac{46}{17}$ & - & $\frac{31}{17}$ & \unit[8.12]{kWh} \\ [1ex]
2GT (Optimized) & $\frac{48}{17}$ &  $\frac{20}{17}$ & $\frac{44}{17}$ & \unit[8.23]{kWh} \\
\end{tabular}
\end{center}
\end{table}

\subsection{Validation}
Whilst validation of our methods using high-fidelity models is subject to future research, we can examine whether our simulation results are in the appropriate range of values.
We compare our simulations and optimal solutions to the baseline design of the 2012 Nissan Leaf, which is equipped with an FGT of $\gamma_1 \approx 1.88$ and $\gamma\_{fd} \approx 4.24$~\citep{Nelson2020}. 
We recreate the transmission with our methodology, and resimulate the vehicle on the CADC (Table~\ref{tb:transmission_designs}) and the New European Driving Cycle (NEDC), since OEM performance data is reported on the latter. 
To make an estimation on the battery energy consumption, the battery and inverter are both modeled using a fixed efficiency, equal to \unit[97.5]{\%} \citep{Steilen2015HydrogenBatteries,HomerEnergy2022} and \unit[95]{\%} \citep{Chang2019ImprovingMOSFETs}, respectively. 
This amounts to a simulated energy consumption over the NEDC of \unit[1.37]{kWh}, which equates to \unit[125.7]{Wh/km}. 
OEM data reports an energy consumption of \unit[126]{Wh/km}: an error of \unit[0.2]{\%}, which leads to the affirmation that our models and methods are sufficiently accurate for this purpose.

\section{Conclusions}\label{sec:conclusions}
In this paper, we aimed at bridging the gap in transmission modeling, between simple models in powertrain sizing and high-fidelity models for component design optimization, in order to rapidly explore the extensive powertrain design space.
To this end, we instantiated a modeling and optimization framework that can simulate different types of transmissions by systematically constructing a configuration, analytically modeling the individual components that it contains, and optimizing the shifting control.
We showcased our framework for a compact family electric car, where the FGT optimum we obtain is slightly different from the baseline design, mainly due to packaging constraints of the final drive within the differential.
Implementing a 2GT can improve the energy consumption, but it might come at the price of higher complexity, volume, mass, and cost.
We validated our framework, showing only a slight deviation in energy consumption when the real-world configuration is modeled, proving that our framework can be used for gear ratio selection, from which expert engineers can further improve the detailed design of the gearbox.

This study introduces the following paths for future research:
First of all, as mentioned previously, our framework could be validated more extensively with high-fidelity models.
Second, the impact of a higher number of gears, possibly combined with more challenging performance requirements, can be readily explored with our framework.
Finally, we can combine our transmission framework with EM and battery sizing, to holistically design an electric vehicle powertrain.

\begin{ack}
We thank Dr. Ilse New and Ir. Fabio Paparella for proofreading this paper.
\end{ack}

\ifextendedversion
	\appendix
	\section{Detailed Gearbox Modeling}
In this appendix, we introduce the more detailed gearbox component models.
Material and type-specific values of the parameters can be found in Table~\ref{tb:appendixparams}, or in the individual sources cited at the introduction of the models.

The width of a gear $b\_g$ is related to its radius through
\begin{equation}
    b\_g = 16\frac{2r\_g}{N\_t}.
\end{equation}
The parameter $\sigma\_s$ used to determine the radius of the shaft $r\_s$ is equal to
\begin{equation}
    \sigma\_s = 2\cdot10^{-2}\left( \frac{2}{5\pi}\right)^{\frac{1}{3}}.
\end{equation}
The parameter $\sigma\_{cl,o}$ used to determine the outer radius of the clutch $r\_{cl,o}$ is equal to
\begin{equation}
    \sigma\_{cl,o} = \left(\frac{3}{2\pi p\_{cl}(N\_z\mu\_{cl})^2(1-\gamma\_{cl}^3)}\right)^{\frac{1}{3}},
\end{equation}
where $p\_{cl}$ is the maximum clutch plate surface pressure, $N\_z$ is the number of clutch plates, and $\mu\_{cl}$ is the friction coefficient of the clutch surface~\citep{Bak2021TorqueConnections}.
The parameter $\sigma\_{cl}$ used to calculate the force required to engage the clutch plate is equal to
\begin{equation}
    \sigma\_{cl} = \pi N\_z \mu\_{cl} p\_{cl}.
\end{equation}

The full expression of the gear sliding losses $F\_{g,sl}$, as a function of the slide-to-roll ratio $S$, is given by
\begin{subequations}
    \begin{equation}
        F\_{g,sl} = e^{f}p^{b\_2}\_h|S|^{b\_3}v^{b\_6}\_e\nu^{b\_7}R^{b\_8}W,
    \end{equation}
    \begin{equation}
        f = b\_1 + b\_4|S|p\_h\log(\nu) + b\_5e^{-|S|p\_h\log(\nu)} + b\_9e^{s},
    \end{equation}
    \label{eq:g_mu_sl}%
\end{subequations}
where $p\_h$ is the Hertzian pressure at the contact surface, $\nu$ is the dynamic viscosity of the oil calculated using the Walther equation~\citep{Seeton2006Viscosity-temperatureLiquids}, $s$ is the surface roughness of the gears, $R$ is the effective radius of curvature, and $b\_1,...,b\_9$ are coefficients determined by the oil~\citep{Xu2007PredictionPairsb}. 

The rolling friction loss is equal to
\begin{equation}
    F\_{g,r} = 4.318 (\tilde{G}\tilde{U})^{0.658}\tilde{Q}^{0.0126}\frac{R}{\alpha},
\end{equation}
where $\tilde{G}$ is a material parameter, $\tilde{U}$ is a speed parameter, $\tilde{Q}$ is a load parameter, and $\alpha$ is the pressure viscosity coefficient of the oil~\citep{Xu2007PredictionPairsb}.

The three churning models are equal to
\begin{subequations}
    \begin{equation}
        P\_{ch,p} = 4\nu b\_g r\_{g,o}^2\omega\_g^2\cos^{-1}{(1-\tilde{h})}
        \label{eq:g_chp}
    \end{equation}
    \begin{multline}
        P\_{ch,f} = \left( b\_{10} \theta\_{oil} \nu^{b\_{11}}\omega\_g^{b\_{12}}r\_{g,o}^{b\_{13}}\right)\\
        \frac{\Big{(}\frac{\pi}{2}{-\sin^{-1}{(1-\tilde{h})}-(1-\tilde{h})\sqrt{\tilde{h}(2-\tilde{h})}\Big{)}}}
        {\sin{\big{(}\cos^{-1}{(1-\tilde{h})}\big{)}}^{b\_{14}}}
    \end{multline}
    \begin{equation}
        P\_{ch,c} = N\_c\nu\delta\_c\omega\_g(r\_{g,o}-r\_{g,i})^2(r\_{g,o}+r\_{g,i})\bigg{(}\frac{D\_3}{r\_{g,o}r\_{g,i}}-D\_4\bigg{)},
    \end{equation}
    \label{eq:g_ch1}%
\end{subequations}
where $\tilde{h}$ is a dimensionless immersion parameter, $\theta\_{oil}$ is the gearbox temperature, $N\_c$ is the amount of cavities that are submerged in oil, $\delta\_c$ is the angle spanning the tooth cavity, and $D\_3$ and $D\_4$ parameters determined by the boundary conditions. 
Constants $b\_{10},...,b\_{14}$ are predetermined constants depending on the flow being laminar or turbulent~\citep{Seetharaman2009}. 

The bearing roll parameter $\sigma\_{b,r}$ is equal to
\begin{equation}
    \sigma\_{b,r} = \phi\_{i}\phi\_{s}G\_{r}\bigg{(}\frac{30 \nu}{\pi}\bigg{)}^{0.6},
    \label{eq:b_rr}
\end{equation}
where $\phi\_{i}$ denotes the oil shear inlet reduction factor, $\phi\_{s}$ is the kinematic starvation factor, and $G\_{r}$ is a constant that is dependent on the bearing type and load~\citep{SKFGroup2008TheMoment}.

The bearing drag parameters $\sigma\_{b,drag}$ and $\bar{\sigma}\_{b,drag}$ are equal to
\begin{equation}
    \sigma\_{b,drag} = 360V\_MKd\_m^5\left(\frac{1}{\pi}\right)^2,
\label{eq:b_drag_a}
\end{equation}
\begin{equation}
    \bar{\sigma}\_{b,drag} = 9.0347\cdot10^{-7}\left(\frac{1}{\pi}\right)^2d\_m^3R\_s\left(\frac{ d\_m^2f\_t}{\pi\nu\_k}\right)^{\epsilon\_{b,drag}},
\end{equation}
where $V\_M$ is the drag loss factor, $K$ is a constant depending on the rolling element type and size, $f\_t$ depends on the oil level and mean bearing diameter $d\_m$, $R\_s$ depends on the inner, outer and mean bearing diameter, oil level and bearing geometry, and $\epsilon\_{b,drag}$ is an exponent defined in Table~\ref{tb:appendixparams}~\citep{SKFGroup2008TheMoment}.

The parameter related to the losses in the synchronizers is equal to
\begin{equation}
    \sigma\_{syn} = \frac{1}{4}\pi\nu\rho\frac{b}{h\_{syn}}d\_c^3,
    \label{eq:syn_2}
\end{equation}
where $b$ is the maximum flow length of the oil, $h\_{syn}$ is the gap between the synchronizer and cone, and $d\_c$ is the diameter of the friction surface~\citep{Dogan2001}.

\begin{table}[]
\begin{center}
\caption{Modeling Parameters}\label{tb:appendixparams}
\begin{tabular}{llll}
Symbol & Variable & Value & Unit \\ \hline
$\sigma_{\mathrm{p}}$ & Pitch radius parameter & 2.5 & [1]\\ 
$\epsilon_{\mathrm{b,drag}}$ & Bearing drag exponent & -1.379 & [1]\\
$\theta\_{oil}$ & Gearbox temperature & 80 & [\unit{$^\circ$C}]\\
$E\_s$ & Young's modulus steel & 190 & [\unit{GPa}]\\
$h$ & Oil level & 0.025 & [\unit{m}]\\
$\phi$ & Gear helix angle & 11 & [$^\circ$]\\
$\nu\_s$ & Poisson ratio steel & 0.3 & [1]\\
\hline
\end{tabular}
\end{center}
\end{table}
	\fi

\bibliography{main,SML_papers,Thijs_de_Mooy,OB_references}             

                                                   







\end{document}